\documentclass[proof]{WileyASNA-v1}
\articletype{Original article}%

\usepackage{makecell}

\received{-}
\revised{-}
\accepted{-}

\begin{document}

\title{A prediction for the 25th solar cycle maximum amplitude}

\author[1]{R. Braj\v{s}a*}
\author[2]{G. Verbanac}
\author[3]{M. Bandi\'{c}}
\author[4]{A. Hanslmeier}
\author[1]{I. Skoki\'{c}}
\author[1]{D. Sudar}

\authormark{R. BRAJ\v{S}A \textsc{et al}}

%\address[1]{\orgdiv{Hvar Observatory}, \orgname{Faculty of Geodesy, University of Zagreb}, \orgaddress{Ka\v{c}i\'{c}eva 26, 10000 Zagreb, \country{Croatia}}}
\address[1]{\orgdiv{Hvar Observatory}, \orgname{Faculty of Geodesy, University of Zagreb}, \orgaddress{Zagreb, \country{Croatia}}}

%\address[2]{\orgdiv{Department of Geophysics}, \orgname{Faculty of Science, University of Zagreb}, \orgaddress{Horvatovac 95, 10000 Zagreb, \country{Croatia}}}
\address[2]{\orgdiv{Department of Geophysics}, \orgname{Faculty of Science, University of Zagreb}, \orgaddress{Zagreb, \country{Croatia}}}

%\address[3]{\orgname{Zagreb Astronomical Observatory},  \orgaddress{Opati\v cka 22, 10000 Zagreb, \country{Croatia}}}
\address[3]{\orgname{Zagreb Astronomical Observatory},  \orgaddress{Zagreb, \country{Croatia}}}

%\address[4]{\orgdiv{Institute for Geophysics, Astrophysics and Meteorology}, \orgname{Institute of Physics, University of Graz}, \orgaddress{Universit\"atsplatz 5, 8010 Graz, \country{Austria}}}
\address[4]{\orgdiv{Institute for Geophysics, Astrophysics and Meteorology}, \orgname{Institute of Physics, University of Graz}, \orgaddress{Graz, \country{Austria}}}

\corres{*R. Braj\v{s}a, \email{roman.brajsa@geof.unizg.hr}}

\presentaddress{Hvar Observatory, Faculty of Geodesy, University of Zagreb, Ka\v{c}i\'{c}eva 26, 10000 Zagreb, Croatia}

\abstract{The minimum - maximum method, belonging to the precursor class of the solar activity 
forecasting methods, is based on a linear relationship between relative sunspot number 
in the minimum and maximum epochs of solar cycles. In the present analysis we apply a
modified version of this method using data not only from the minimum year, but also
from a couple of years before and after the minimum. The revised 13-month
smoothed monthly total sunspot number data set from SILSO/SIDC is used. 
Using data for solar cycle nos. 1-24 the largest correlation coefficient ($CC$) is
obtained when correlating activity level 3 years before solar cycle minimum with the
subsequent maximum ($CC = 0.82$), independent of inclusion or
exclusion of the solar cycle no. 19. For the next solar maximum of the cycle
no. 25 we predict: $R_{\rm max} =121 \pm 33$.
Our results indicate that the next solar maximum (of the cycle no. 25) will be of the
similar amplitude as the previous one, or even something lower. This is in accordance
with the general middle-term lowering of the solar activity after the secular maximum
in the 20th century and consistent with the Gleissberg period of the solar activity.
The reliability of the "3 years before the minimum" predictor is experimentally
justified by the largest correlation coefficient and verified with the Student t-test. It is satisfactorily explained with the
two empirical well-known findings: the extended solar cycle and the Waldmeier effect.
Finally, we successfully reproduced the maxima of the last four solar cycles, nos. 21-25, using the 3 years before the minimum method.
}


\jnlcitation{\cname{%
\author{Braj\v{s}a R.}, 
\author{G. Verbanac}, 
\author{M. Bandi\'c}, 
\author{A. Hanslmeier},
\author{I. Skoki\'c}, and 
\author{D. Sudar}} (\cyear{2021}), 
\ctitle{A prediction for the 25th solar cycle maximum amplitude}, \cjournal{Astron. Nachr.}, \cvol{2021;00:1--7}.}

%\fundingInfo{Croatian Science Foundation MSOC-7549; 
%Austrian-Croatian Bilateral Scientific Project "Comparison of ALMA observations with MHD-simulations of coronal waves interacting with coronal holes".}

\maketitle

\section{Introduction}

"Prediction is difficult, especially of the future." This quote is attributed to Niels Bohr 
\citep{Peitgen2004}. In this work we make a prediction of the amplitude of the next solar cycle maximum, knowing 
that the minimum between solar cycles no. 24 and no. 25 was in December 2019.

The solar magnetic activity cycle belongs to the few problems in 
contemporary solar physics which are not resolved to the satisfactory level. 
Besides their important place within solar physics \citep{Harvey1992, Hoyng1992, Wilson1994, Ossendrijver2003, Rudiger2004, 
Thomas2008, Charbonneau2013, Charbonneau2014, Charbonneau2020, Miesch2016, Brun2017}, 
they also have a practical role as drivers of the space weather \citep{Koskinen2017} and possible causes of the climatic change 
\citep{Rapp2008, Gray2010}. 

Moreover, solar cycle forecasting 
serves as a very good tool to estimate the upcoming geomagnetic activity, which is very important for planning the satellite missions 
within the Earth's magnetosphere, orbital correction of the satellites already in the orbits around the Earth, and protection of on board 
instruments that monitor the ionospheric-magnetospheric state.
\citet{Verbanac2011} performed a thorough analysis of the relationship between various solar and geomagnetic activity 
indices on 1-year data resolution.
Based on their work, which demonstrated quite strong relationship between all considered geomagnetic and solar activity parameters, 
the 1-year prediction of the level of geomagnetic disturbances is possible.
Namely, they found the average 1-year time delay of the geomagnetic activity (quantified by geomagnetic Ap index) behind all solar indices 
including sunspot numbers used in the present study.
Further, the delay of Ap index for two years with respect to 10.7 cm flux is presented in \citet{Verbanac2010}.
Based on that it is clear that the prediction of the magnitude of the solar cycle is very useful in estimating the geomagnetic activity 
at least one year in advance.

There is also an additional, "purely astronomical" motivation for the precise monitoring of the solar activity and for developing 
reliable solar cycle prediction tools. It is widely known that the sky brightness is well correlated with the solar activity, in particular with 
the 10.7 cm flux. Solar EUV radiation, variable during the solar activity cycle, influences the airglow in the upper Earth's atmosphere, 
increasing the sky brightness during the maximum of activity \citep{Walker1988}. 
The effect should be taken into account for efficient planning 
of astronomical observations with large facilities, such as Rubin/LSST\footnote{\url{https://www.lsst.org/}}
 \citep{Ivezic2019}
and ESO\footnote{\url{https://www.eso.org/public/teles-instr/}} 
telescopes \citep{Leinert1995, Patat2003}.
Moreover, solar cycle prediction is also important for optimal observational plan of those solar phenomena which are strongly correlated 
with the solar activity, such as active regions, sunspots, and flares. This is again very important for large facilities, 
such as ALMA\footnote{\url{http://www.almaobservatory.org}}, 
where the observing time is sparse and spread over almost all types of astronomical objects, with solar observations having only a small 
fraction of the whole observing time \citep{Brajsa2018}. 

The most common index of solar activity is the sunspot number. 
After several hints and alerts that the sunspot number series in use needs to be revised for several inconsistencies, 
a serious program for recalibration of the sunspot number was started 
\citep{Cliver2013, Clette2014, Cliver2015} and successfully finished in the mid of 2015 \citep{Clette2016}. 
There are several reasons why the "old" sunspot number should be corrected, the most important being 
(i) mutually inconsistent sunspot number series and group sunspot number series, (ii) more and more evidence that the 
sunspot number series is not homogenous showing important discontinuities obviously not representing real changes 
of the solar activity, and (iii) a curious secular trend in 
solar activity inferred from variations of the sunspot number \citep{Cliver2013, Clette2014}. The recalibration process had 
to be done very carefully, since the sunspot number is widely used in studies of the solar dynamo, terrestrial climate 
change and space climate change \citep{Cliver2015}. 
Finally, the "new", improved sunspot number\footnote{\url{http://sidc.oma.be/silso/newdataset}} was officially introduced on July 1st, 2015. 
 
There are two general types of methods for the solar cycle predictions in use: 
the empirical methods and the methods relying on MHD dynamo models \citep{Hathaway1999, Hathaway2009, Hathaway2015, 
Brajsa2009, Petrovay2010, Petrovay2020, Pesnell2012}. 
The empirical methods are further divided into the extrapolation and the precursor methods. 

In present work we use the empirical approach, a modified minimum - maximum method, which belongs to the precursor class 
of methods. It was introduced by \citet{Wilson1990a} and the input data include the sunspot number and various 
geomagnetic indices. 
Although the method is relatively simple, up to now it has not been used as much as it would be expected 
\citep{Brajsa2009, Ramesh2012, Pishkalo2014, Brajsa2015}.
Basically the method uses knowledge of some solar parameters or proxies in and around solar minimum to 
predict the level of activity in the next solar maximum. The solar parameter used in present work is the sunspot 
number. The modification used here consists of using the input data shifted in time from the minimum epoch. 

Present work has three main aims. (i) To check whether the assumption that 3 years before the activity minimum is the best epoch 
to predict the next solar maximum is true. This assumption was made independently by \citet{Svalgaard2005} and by 
\citet{Cameron2007} using different methods for solar cycle forecasting than in present work. Their methods and arguments will 
be discussed in more detail later in this paper. The second aim of this analysis is (ii) To 
make a prediction of the amplitude of the next solar cycle maximum, using the modified minimum - maximum method, 
taking into account the previous assumption, and knowing 
that the minimum between solar cycles no. 24 and no. 25 was in December 2019. 
Finally,  (iii) we check the reliability of the method by reproducing the maxima of the last four solar cycles, nos. 21-25, 
using the proposed modified minimum - maximum method.

\section{Data set and reduction method}

\begin{center}
\begin{table*}[th!]
\centering
\caption{The solar cycle number, the epochs of solar minima and maxima ($T_{min}$ and $T_{max}$, respectively) and the extreme values of 
the monthly smoothed sunspot numbers in corresponding months ($R_{min}$ and $R_{max}$, respectively).\label{table_1}}  
\begin{tabular}{c c c c c r r r}
\toprule
{Cycle No.} & {$T_{min}$ (year)} & {(month)} & {$T_{max}$ (year)} & {(month)} & {$R_{min-3}$} & {$R_{min}$} & {$R_{max}$} \\ 
\hline
1               &   1755                 &  3           &   1761                   &  6          &  75.5  &   14.0         &   144.1 \\ 
2               &   1766                 &  6           &   1769                   &  9          &  76.3  &   18.6         &   193.0 \\ 
3               &   1775                 &  6           &   1778                   &  5          & 113.0    &   12.0         &   264.3 \\ 
4               &   1784                 &  9           &   1788                   &  2          & 101.0    &   15.9         &   235.3 \\ 
5               &   1798                 &  4           &   1805                   &  2          &  46.6  &     5.3         &     82.0 \\ 
6               &   1810                 &  7           &   1816                   &  5          &  15.9  &     0.0         &     81.2 \\ 
7               &   1823                 &  5           &   1829                   & 11          &  30.2  &     0.2         &   119.2 \\ 
8               &   1833                 & 11          &   1837                   &  3          &  106.6 &   12.2         &   244.9 \\ 
9               &   1843                 &  7           &   1848                   &  2          & 103.8  &   17.6         &   219.9 \\ 
10             &   1855                 & 12          &   1860                   &  2          &  84.5  &     6.0         &   186.2 \\ 
11             &   1867                 &   3          &   1870                   &   8         &  88.5  &     9.9         &   234.0 \\ 
12             &   1878                 & 12          &   1883                   & 12          &  20.9  &     3.7         &   124.4 \\ 
13 		&   1890 		     &   3 	      &   1894                   &  1          &  21.0  &     8.3         &   146.5 \\ 
14 		&   1902 		     &   1 	      &   1906  		     &  2          			& 34.0     &     4.5         &   107.1 \\ 
15 		&   1913                 &   7          &   1917                   &  8          &  29.4  &     2.5         &   175.7 \\ 
16             &   1923                 &   8          &   1928                   &  4          &  58.2  &     9.4         &   130.2 \\ 
17 		&   1933 	             &   9           &   1937                   &  4          &  51.2  &     5.8         &   198.6 \\ 
18 		&   1944                &   2           &   1947                   &  5          &  91.2  &   12.9         &   218.7 \\ 
19 		&   1954                &   4           &   1958                   &  3          &  100.2 &     5.1         &   285.0 \\ 
20             &   1964 	             & 10           &   1968                   & 11         &  73.0   &   14.3         &   156.6 \\ 
21 		&   1976                &   3           &   1979                   & 12         &  62.8   &   17.8         &   232.9 \\ 
22 		&   1986 	             &   9           &   1989                   & 11         &  91.7   &   13.5         &   212.5 \\ 
23 		&   1996 	             &   8 	      &   2001  	             & 11  	    &   73.6   &   11.2         &   180.3 \\ 
24 		&   2008 	             & 12 	      &   2014  	             &  4   	    &  36.0    &    2.2          &   116.4 \\ 
25		&   2019		    &	12	      &   --     		    & -- 	    &  28.5  &    1.8		&    --      \\
\bottomrule
\end{tabular}
\end{table*}
\end{center}

We use the 13-month smoothed monthly total sunspot number data set for the period 1749 to March 2020 (which became 
available in October 2020), from the Sunspot Index and 
Long-term Solar Observations (SILSO\footnote{\url{http://sidc.be/silso/datafiles}}) Data Center of the ROB, Brussels
\citep{sidc}. 
The data from the solar cycle no. 1 up to minimum between solar cycles no. 24 and no. 25 are presented in Table \ref{table_1}. 

A modified version of the minimum - maximum method is used considering not only the minimum activity
years, but also years before and after the epoch of the solar activity minimum. 
We use the input data for solar cycles nos. 1--24 from Table \ref{table_1} and data of the epochs of minima from SILSO. 
The procedure is repeated excluding the data for solar cycle no. 19. 
The solar cycle no. 19 is a rather unusual cycle in which the highest measured activity maximum was preceded by a relatively 
low minimum (Table \ref{table_1}). There are many indications that this solar cycle was not a typical one, even maybe a real 
outlier \citep{Wilson1990b, Temmer2006}. 

\begin{figure}
\centering
\includegraphics[width=\linewidth]{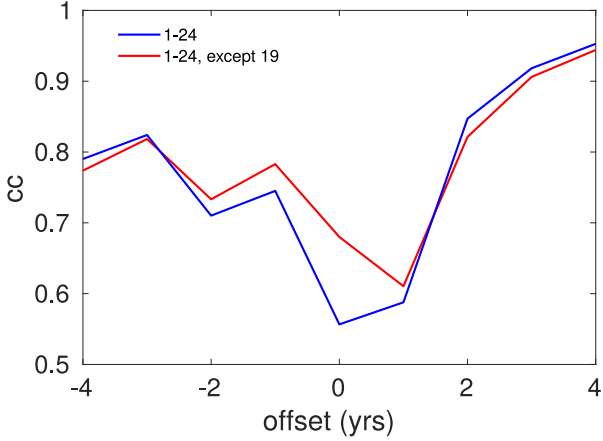}
\caption{The correlation coefficient $CC$ of the minimum $-$ maximum relationship for different values of
the time offset in years. Solar cycles nos. 1-24 were analyzed using monthly smoothed sunspot numbers. 
Results obtained with and without solar cycle no. 19 are presented with different colors, as indicated in the 
legend.}
\label{Fig_1}
\end{figure}

The correlation coefficient, $CC$, is investigated as a function of the time offset in years (Figure~\ref{Fig_1}).
A maximum in the $CC$ 3 years before the minimum is clearly seen in Figure~\ref{Fig_1}. 
After the minimum the $CC$ sharply rises as the cycle moves toward the maximum. 
The two datasets converge and therefore the correlation approaches 1, $CC = 1$.
It is interesting that the $CC$ has the lowest value in the epoch of the solar minimum. 
A general behavior of the $CC$ as a function of the offset without cycle no. 19 is similar to the 
case with all solar cycles, with the exception that the minimum in the $CC$ value occurs one year after the 
activity minimum. 

\begin{figure}
\centering
\includegraphics[width=\linewidth]{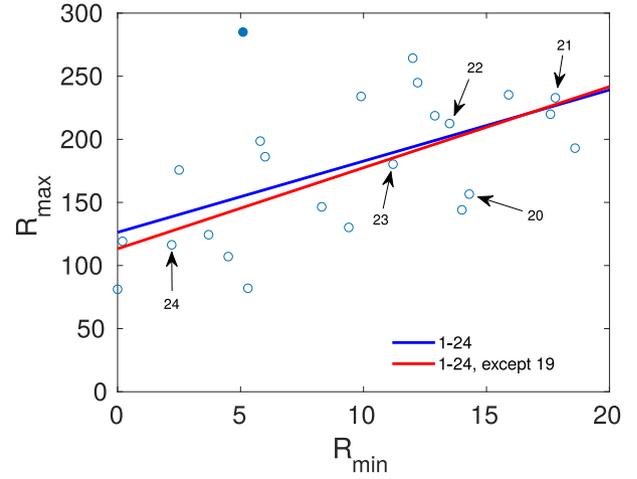}\par \caption{The peak smoothed monthly sunspot number in solar cycle maxima as a function
of the same quantity in the preceding solar minimum, for solar
cycles nos. 1-24.
Least-square fits obtained with and without solar cycle no. 19 are presented with different colors, as indicated in the 
legend. The value for the solar cycle no. 19 is represented with the filled circle, while all other data points are 
represented with open circles.}
\label{Fig_2}
\end{figure}

\begin{figure}
\centering
  \includegraphics[width=\linewidth]{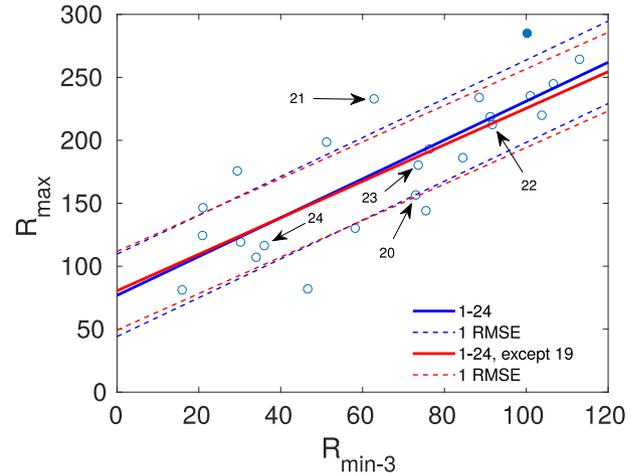}\par \caption{Similar to Figure~\ref{Fig_2}, but for the values 3 years before the activity minima. 
Dashed lines represent uncertainty of the fits.}
    \label{Fig_3}
\end{figure}

In Figure~\ref{Fig_2} the linear least-square fit  for $R_{\rm max}$  vs.  $R_{\rm min}$   is presented and in Figure~\ref{Fig_3} the similar fit 
for $R_{\rm max}$ vs.   $R_{\rm min - 3}$, together with the 1$\sigma$ uncertainty boundaries. 
$R_{\rm min - 3}$ represents the smoothed monthly sunspot number 3 years before the activity minimum. Both procedures were 
performed with and without the unusual solar cycle no. 19. 
The exclusion of the cycle no. 19 gives slightly different fits (Figures~\ref{Fig_2} and \ref{Fig_3}). 

\section{Results}

We present the least-square fit parameters of the general form: 
\begin{equation}
\label{Eq_1}
R_{\rm max} = a \cdot R_{\rm min}(t) + b
\end{equation}

\noindent 
where $R_{\rm max}$ is the amplitude of the cycle maximum and $R_{\rm min}$ is the activity value in the minimum or before. 
Taking into account input data for solar cycles 1 -- 24 we get: 
\begin{align} 
\label{Eq_2}  
   R_{\rm max}=&(5.6 \pm 1.8) \cdot R_{\rm min} + (126 \pm 20), &CC = 0.56  \\
\label{Eq_3}   
   R_{\rm max}=&(1.5 \pm 0.2) \cdot R_{\rm min - 3} + (77 \pm 17), &CC = 0.82 
\end{align}

\noindent		
We now repeat the formulae obtained without the solar cycle no. 19: 
\begin{align}
\label{Eq_4}
R_{\rm max} =& (6.4 \pm 1.5) \cdot R_{\rm min} + (113 \pm 17), &CC = 0.68
\\
\label{Eq_5}
R_{\rm max} =& (1.4 \pm 0.2) \cdot R_{\rm min - 3} + (80 \pm 16), &CC = 0.82
\end{align}
			
Taking into account the date of the current minimum (December 2019) we can use the observed monthly smoothed value $R_{\rm min} = 1.8$
and the value three years before (December 2016) $R_{\rm min - 3} = 28.5$
to calculate the maximum of the current 25th solar cycle.
For the four cases  described earlier, the following predictions for the maximal amplitude 
of the solar cycle no. 25 are calculated using Equations (\ref{Eq_2}) -- (\ref{Eq_5}), respectively: 
\begin{align}
\label{Eq_6}
R_{\rm max} =& 137 \pm 48 \hskip 0.2 cm (R_{\rm min} {\rm ,} \, R_{\rm max}, \rm nos. \, 1-24)
\\
\label{Eq_7}
R_{\rm max} =& 121 \pm 33 \hskip 0.2 cm (R_{\rm min-3} {\rm ,} \, R_{\rm max}, \rm nos. \, 1-24)
\\
\label{Eq_8}
R_{\rm max} =& 125 \pm 40 \hskip 0.2 cm (R_{\rm min} {\rm ,} \, R_{\rm max}, \rm excl. \, no. \, 19)
\\
\label{Eq_9}
R_{\rm max} =& 122 \pm 31 \hskip 0.2 cm (R_{\rm min-3} {\rm ,} \, R_{\rm max}, \rm excl. \, no. \, 19)
\end{align}

\noindent The given errors represent the RMSE. We see that excluding the solar cycle no. 19 narrows the prediction for the 
two subcases, minimum vs. maximum and minimum - 3 years vs. maximum. 

To further investigate the predictive reliability of both approaches, we applied them to predict the maximum amplitude of the last four completed solar cycles nos. 21--24 and compare the result with the actual observed amplitude. As before, we considered cases with and without the solar cycle 19. For each of the four cycles tested, the linear fit coefficients were calculated by using data only from the cycles older than the one being predicted. Although this means that the number of data points available for fitting is decreasing for each past cycle, doing it this way better simulates actual past predictions when knowledge of future cycles was not available. This method also gives the stability of the fit coefficients during the last several cycles.

The results of the maximum amplitude prediction for the last four solar cycles, with addition of the next 25th cycle, and comparison with the actual measured amplitude are given in Table~\ref{table_2}  and graphically presented in Figure~\ref{Fig_4} which is given in the Appendix. 
Upper two table sections show results from the minimum--maximum method, while lower two sections list values obtained using the 3 years before minimum method. It can be seen that the slope coefficient $a$ remains pretty stable with its error generally slightly decreasing when more cycles are used. The same can be concluded for the intercept coefficient $b$, when the $R_{\rm min}$ vs $R_{\rm max}$ method is used, except in the last case when all 24 cycles are used and the value of $b$ slightly decreases but is still fairly within 1$\sigma$. This is due to the very low minimum of the 24th cycle. When 3 years before the minimum method is used, $b$ value again shows statistically insignificant variation.

On the other hand, the CC slowly increases in value when more and more cycles are used, for the minimum - maximum method. The only exception is cycle 21 (when included in the analysis) which worsens the correlation slightly for the 3 years before minimum method. As can be seen in Figure~\ref{Fig_3}, cycle 21 falls far from the best fit line, more the 1$\sigma$ away, thus decreasing the overall correlation. This is also the reason behind the 2$\sigma$ difference in the predicted and observed values of the cycle 21. Furthermore, it is worth mentioning that the removal of the cycle 19 improves the RMSE of the minimum--maximum method but has no significant effect on the 3 years before the minimum method.

In summary, 3 years before the minimum method is preferred over the minimum--maximum method since it has better CC and lower statistical errors. This is also visible when comparing Figures \ref{Fig_2} and \ref{Fig_3}. Data points are grouped more closer to the regression line in case of the 3 years before the minimum method. However, for a particular cycle it may happen (e.g. cycle 21) that a point is further away from the regression line for $R_{\rm min-3}$ than for $R_{\rm min}$. In that case, $R_{\rm min}$ would give a better prediction, but it is a random event which depends on the particular cycle.

\section{Discussion}

\begin{center}
\begin{table*}[t]
\centering
\caption{Comparison of the predicted and observed solar cycle maxima amplitudes for cycles 21-25. The column named "cycles used" denotes how many previous solar cycles were employed to calculate the fit coefficients ($a$, $b$) and their errors ($\sigma_a$, $\sigma_b$), as well as correlation coefficients (CC) and predicted amplitudes ($R_{\rm max}$) of the next cycles based on the observed $R_{\rm min}/R_{\rm min-3}$ of the same cycles.\label{table_2}}  
\begin{tabular}{c c r r r r r c r c r c}
\toprule
 
{Method} & 
 \multirow{2}{*}{\makecell{ Cycles \\  used}} &  {$a$} & 
{$\sigma_a$} & {$b$} & {$\sigma_b$} & {CC} & 
   \multirow{2}{*}{\makecell{ Next \\  cycle}} & 
  {$R_{\rm min}$} &
 \multirow{2}{*}{\makecell{ Predicted \\  $R_{\rm max}$ }} &
 {RMSE} &  
 \multirow{2}{*}{\makecell{ Observed \\  $R_{\rm max}$ }}
 \\ 

\cmidrule(lr){9-9} 
 
 & & & & & & & & {$R_{\rm min-3}$} &  & & \\
 
\hline

& 1-24 &  5.6 & 1.8 & 126.3 & 19.9 & 0.56 & 25 & 1.8 & 136.5 & 48.0 & - \\
$R_{\rm min}$ vs $R_{\rm max}$ & 1-23 &  5.4 & 1.9 & 129.6 & 21.5 & 0.53 & 24 & 2.2 & 141.4 & 48.8 & 116.4 \\
with & 1-22  & 5.4 & 2.0 & 129.8 & 22.0 & 0.53 & 23 & 11.2 & 190.6 & 49.9 & 180.3 \\
cycle 19 & 1-21 &  5.4 & 2.1 & 129.9 & 22.6 & 0.52 & 22 & 13.5 & 202.4 & 51.1 & 212.5 \\
 & 1-20 &  5.3 & 2.3 & 130.5 & 23.6 & 0.49 & 21 & 17.8 & 224.1 & 52.4 & 232.9 \\

\hline
& 1-24 & 6.4 & 1.5 & 113.3 & 17.0 & 0.68 & 25 & 1.8 & 124.8 & 39.8 & - \\
$R_{\rm min}$ vs $R_{\rm max}$& 1-23 & 6.3 & 1.6 & 115.0 & 18.6 & 0.66 & 24 & 2.2 & 128.8 & 40.7 & 116.4 \\
without & 1-22 & 6.3 & 1.7 & 115.1 & 19.1 & 0.66 & 23 & 11.2 & 185.8 & 41.7 & 180.3 \\
cycle 19 & 1-21 & 6.2 & 1.8 & 115.2 & 19.6 & 0.65 & 22 & 13.5 & 199.4 & 42.7 & 212.5 \\
& 1-20 & 6.1 & 1.9 & 115.8 & 20.5 & 0.62 & 21 & 17.8 & 225.0 & 43.8 & 232.9 \\

\hline
& 1-24 & 1.5 & 0.2 & 76.8 & 16.8 & 0.82 & 25 & 28.5 & 120.8 & 32.7 & --  \\
$R_{\rm min-3}$ vs $R_{\rm max}$ & 1-23 & 1.5 & 0.2 & 79.1 & 17.8 & 0.82 & 24 & 36.0 & 133.8 & 33.3 & 116.4 \\
with & 1-22 & 1.5 & 0.2 & 79.4 & 18.2 & 0.82 & 23 & 73.6 & 191.4 & 34.0 & 180.3 \\
cylce 19 & 1-21 & 1.5 & 0.3 & 79.2 & 18.7 & 0.81 & 22 & 91.7 & 219.5 & 34.8 & 212.5 \\
& 1-20 & 1.5 & 0.2 & 75.7 & 18.0 & 0.84 & 21 & 62.8 & 172.3 & 33.0 & 232.9 \\

\hline

& 1-24 & 1.4 & 0.2 & 80.5 & 16.2 & 0.82 & 25 & 28.5 & 121.8 & 31.2 & -- \\
$R_{\rm min-3}$ vs $R_{\rm max}$ & 1-23 & 1.4 & 0.2 & 82.8 & 17.1 & 0.81 & 24 & 36.0 & 134.1 & 31.7 & 116.4 \\
without & 1-22 & 1.4 & 0.2 & 83.0 & 17.6 & 0.81 & 23 & 73.6 & 188.1 & 32.5 & 180.3 \\
cycle 19 & 1-21 & 1.4 & 0.3 & 82.9 & 18.2 & 0.81 & 22 & 91.7 & 214.1 & 33.3 & 212.5 \\
& 1-20 & 1.4 & 0.2 & 79.5 & 17.1 & 0.83 & 21 & 62.8 & 169.6 & 31.0 & 232.9 \\

\bottomrule
\end{tabular}
\end{table*}
\end{center}

First, we can confirm the assumption that 3 years before solar activity minimum is the best time when reliable prediction for the 
next maximum can be made. This conclusion is supported by the highest correlation coefficient at that epoch (Figure~\ref{Fig_1}).
So, the assumption of the importance of the time 3 years before minimum, made by \citet{Svalgaard2005} and 
\citet{Cameron2007}, is here independently confirmed and reaffirmed. Moreover, the curves for the two cases, with and without 
solar cycle no. 19, have almost the same value at the epoch 3 years before solar minimum (Figure~\ref{Fig_1}). An important 
implication of this fact is that excluding the solar cycle no. 19 does not have a significant influence on the predictive skill of the 
method, if the modified procedure $R_{\rm max}$ vs. $R_{\rm min - 3}$ is considered. Finally, we can also easily understand 
the minimal value of the correlation coefficient for the $R_{\rm max}$  vs.  $R_{\rm min}$ case when solar cycle no. 19 is included
(Figure \ref{Fig_1}). 
The explanation is based on the fact that the highest solar activity maximum observed in solar cycle no. 19 was preceded by a 
relatively low minimum (Table~\ref{table_1}). 
This blurs the correlation, but the influence is completely removed when data 3 years before the minimum 
are considered. 

A question can be raised if the difference between the minimum vs 3 years before the minimum CCs is statistically significant. To investigate this, we performed an unequal variances $t$--test \citep{student1908, welch1947, Press2002, Ivezic2014} on two samples corresponding to two prediction methods. T-test is used to determine if two populations have equal means, i.e. are the means of two data sets significantly different from each other. This is numerically characterized by a $t$-value, a distance between the two means in terms of standard deviations, and a $p$-value, a probability of obtaining the observed, or more extreme, value when the null hypothesis is true. The null hypothesis, in our case, is that both population means are equal. For the $R_{\rm min}$ and $R_{\rm min-3}$ data sets, using all 24 cycles, the null hypothesis is strongly rejected by the $t$--test with a $p$-value well below the usual 0.05 threshold ($1.5\times10^{-11}$). We also compared $R_{\rm min-3}$ and $R_{\rm min-t}$, where $t$ is time in years from the minimum epoch, for $t$ values of 1 to 4 years. As expected, the $t$-test gave the largest $p$-values for $R_{\rm min-3}$ neighboring data sets ($t$ equal to 2 and 4 years, corresponding $p$-values are 0.0026 and 0.0019, respectively), and increasingly smaller values for more distant data sets. From the analysis above, we can conclude that $R_{\rm min-3}$ and $R_{\rm min}$ data sets used for the prediction of the amplitude of the next maximum are statistically different and do not represent two samples of the same population.

Considering 3 years before the minimum as the best indicator, our prediction for the next solar maximum is  
$R_{\rm max} =121 \pm 33$.
A very similar result is obtained when solar cycle no. 19 is excluded: 
$R_{\rm max} = 122 \pm 31$.
If we repeat the procedure using data from solar cycles nos. 1-23 (up to the year 2008) to predict $R_{\rm max}$ 
of the solar cycle no. 24, we obtain $R_{\rm max} = 134 \pm 33$. 
However, the actual value was 116.4 (Table \ref{table_2}).
We emphasize that we should not directly compare our prediction of 121 with the value 116.4 and make a conclusion about the fact which solar 
cycle is or will be stronger.
It is important to consider the RMSE values (note that the magnitude of the RMSE for both predicted $R_{\rm max}$ values are the same) 
which give the lower and upper limits for the predicted value.
Our method predicts for solar cycle no. 24 the $R_{\rm max} = 134 \pm 33$
and for solar cycle no. 25 the $R_{\rm max} =121 \pm 33$. 
Thus for the solar cycle no. 24 the predicted values are in the range 101--167 and for the solar cycle no. 25 in the range 88--154.
Note that differences between predicted values of solar cycle no. 24 and solar cycle no. 25 are not a consequence of the 
fitting procedure (at least the effect is not significant).
Namely, with 24 points (see Figure~\ref{Fig_3}) we obtained Equation (\ref{Eq_7}), 
and with 23 points (the cycle which we want to predict, no. 24, is omitted) the obtained expression is similar. 
The main reason for the difference are the values of $R_{\rm min - 3}$. In the prediction of the solar cycle no. 24 the value 36.0 for the epoch 
12/2005 was used (3 years before the minimum of the cycle no. 24 which was at the epoch 12/2008), whereas in the prediction of solar cycle 25 the value 28.5 for the epoch 12/2016 was employed (3 years before the minimum of the cycle no. 25 which was at the epoch 12/2019), 
 see also Table~\ref{table_1}.
The used $R_{\rm min - 3}$ input value for predicting the solar cycle no. 25 is lower and consequently also the calculated lower and upper 
limits, which indicate the possibility that the upcoming maximum will be lower.
This is probably in accordance with the general middle-term lowering of the solar activity after the secular maximum in the 20th century and consistent with the Gleissberg period (the time scale of about a century) of the solar activity.
However, the indication of lower upcoming minimum must be taken with some caution, as one should compare the trend of prediction for many cycles (not only cycle no. 24 and no. 25) with the real values to see if the method is capable to track the real trend, as was done in present work.

We now compare our prediction results with some early predictions found in the literature. 
So, \citet{Du2020} applied the precursor method using the preceding minimum of $aa$ geomagnetic index and forecasted the maximum of the 
solar cycle no. 25 to be $R_{\rm max} = 151 \pm 17$, which is about 30\% larger than the previous maximum, and also larger than our 
prediction, but still within 1$\sigma$ error on both sides. 
On the other hand, \citet{Miao2020} predicted $R_{\rm max} = 122 \pm 33$ for the maximum of the cycle no. 25 using a combination of the 
Ohl's prediction method and minimum $aa$ geomagnetic index. This forecasted value is very close to our predicted value (Equation \ref{Eq_7}).

It is worth to note that \citet{Petrovay2020} gives the formulae for the minimum -- maximum method based on solar cycles nos. 1--24, 
excluding cycle no. 19, for the cases 
$R_{\rm max}$  vs.  $R_{\rm min}$ and $R_{\rm max}$ vs.   $R_{\rm min - 3}$. 
The formulae are given in the review of \citet{Petrovay2020} as Equations (10) and (11), which correspond to our Equations (\ref{Eq_4}) and (\ref{Eq_5}). We note, 
however, that the two sets of formulae are similar but not equal and that \citet{Petrovay2020} does not provide the errors of the 
linear least-square fit parameters, which are calculated and used in present work. Based on the data available at the time of writing the 
review, \citet{Petrovay2020} 
obtained that the maximal possible value for the next solar maximum is $R_{\rm max} = 147$, which is now put to the lower 
values. 

We can also raise the question why the predictor 3 years before the minimum gives the highest correlation coefficient $CC$, implying the 
most reliable epoch for the prediction, and to understand the importance of the value 3 years before the minimum which will determine 
the maximum of the next cycle. 
However, before discussing in detail the two papers \citep{Svalgaard2005, Cameron2007} 
and the lines of reasoning of their authors, we briefly repeat 
some important ingredients of the self-exciting oscillating dynamo model of the Babcock-Leighton type. 
The model, along with later modifications and improvements, assumes that the differential rotation winds up the large-scale 
dipolar global poloidal field.
This global poloidal magnetic field prevails  the large-scale distribution of the surface field around solar activity 
minimum. The wound up poloidal field produces subsurface toroidal field which later moves across the surface and 
manifests itself as the sunspot activity of the next cycle. The strength of the polar magnetic field during the declining 
phase of one solar cycle is considered to be a sign of the highest amplitude of the solar activity in the next cycle.
It is important to point out that this process takes place in the layers just below the solar surface.
So, it can be directly observed during the time embracing the previous cycle. Consequently it is plausible to assume that the 
maximal value of the reversed polar magnetic field produced after the solar maximum will be a good precursor for the 
amplitude of the poloidal field from which the next toroidal field will be produced by the differential rotation.
So, the precursor method based on the polar field appears to be satisfactorily rooted in solar physics. 

\citet{Svalgaard2005} made a solar maximum prediction based on the correlation between the amplitude of the solar 
magnetic dipole moment at the time 3 years before the minimum of activity with the amplitude of the next solar 
maximum. During the activity minimum the polar magnetic field attains maximal values and it changes polarity during the 
maximum of activity.
The cause of this magnetic field reversal is the motion of unipolar magnetic flux from lower/medium latitudes towards 
the poles. This new flux cancels the flux of the opposite polarity already present there and the new magnetic field of the 
opposite polarity is eventually generated at the poles \citep{Wang1989}. 
The new activity begins to destroy the polar magnetic field yielding the phenomenon of the strongest polar field 
during the time of approximately 3 years before the cycle minimum. Using the average magnetic field value 
during this time interval as the precursor for the next solar cycle, \citet{Svalgaard2005} succeeded to predict the maximal amplitude of the 24th solar cycle rather well, although the predicted value was lower by 10\% compared to the 
really measured value. 

\citet{Cameron2007} investigated  efficiency and reliability of solar cycle forecasting testing various methods based on precursors 
and magnetic flux transport models. 
As a predictor they used the magnetic flux protruding over the solar equator. This is justified by the fact that in the 
Babcock-Leighton dynamo model this quantity is related to the global dipole magnetic field from which the toroidal 
field for the next solar cycle is produced.
As proxies for solar activity they used several measured and calculated quantities, such as sunspot number, sunspot area, 
polar field, equator flux and dipole component. These authors have concluded that the activity level 3 years before sunspot 
minimum is the best precursor for prediction of the amplitude of the subsequent maximum. 

\citet{Cameron2007} also offered an explanation about the origin of the predictive skill of the methods they used. 
The authors claim that 
to understand the predictive ability of the activity level in the descending phase of the activity cycle it is not needed 
to establish a physical relation between the surface magnetic manifestations in the previous and subsequent solar 
cycles. It is enough to embrace the two well-known properties of the sunspot number series. The first one is the concept 
of extended solar cycle \citep{Harvey1992}. This concept emphasizes the observational fact of the simultaneous 
appearance of sunspots at high latitudes, belonging to the new cycle, and at low latitudes, belonging to the 
previous, declining cycle. The second property is the Waldmeier effect \citep{Waldmeier1935, Brajsa2009} which 
relates the ascending time of a cycle toward its maximal phase and the highest amplitude of the maximum. The 
Waldmeier effect means that stronger cycles rise faster towards their maxima. 
A combination of these two effects results in a systematic temporal shift of the minimum between the two 
subsequent cycles with various amplitudes, when using the activity indices which are averaged over latitudes 
(e.g., the sunspot number or the sunspot area). The final outcome of these two effects is the earlier occurrence 
of the minimum if the next cycle is stronger than the previous one, and the later occurrence of the minimum for 
the opposite case (a weaker following cycle). So, a higher level of activity 3 years before the minimum is a  
logical and necessary consequence of a statistically earlier epoch of the minimum. This line of reasoning can also 
help to understand yet another observational finding, namely the fact that the stronger cycles are statistically 
mostly preceded by the shorter cycles \citep{Hathaway1994, Hathaway1999, Hathaway2002}.   
So, it is not needed to search for a physical mechanism which would relate surface phenomena observed in subsequent 
solar cycles to justify the prediction using the precursors in the decreasing phase of activity. 
\citet{Cameron2007} just offered a simple empirical explanation for the very useful "3 years before the minimum" predictor. 

Finally, we note that the main problem in the reliability of the solar activity forecasting is the influence of the non-linear effects in the solar dynamo, 
which plays the major role in establishing and maintaining solar activity cycle \citep{Hanslmeier2013}.  
According to the present knowledge, solar activity cycle and the underlying solar MHD dynamo show properties on the edge of 
a chaotic process \citep{Hanslmeier2020}. So, the predictability is limited and the non-linear effects are the main source of uncertainty 
of any prediction. 

\section{Summary and conclusions}

Our prediction for the  maximum of the cycle no. 25 gives $R_{\rm max} =121 \pm 33$. 
This result is based on the modified minimum - maximum method, correlating the level of activity 3 years before the minimum 
with the subsequent maximum. For this procedure the correlation coefficient is highest and inclusion or exclusion of the somewhat special 
cycle no. 19 does not influence the predictive skill. 
The reliability of the "3 years before the minimum" predictor is experimentally justified by the largest correlation coefficient and sufficiently 
explained with the two empirical well-known findings: the extended solar cycle and the Waldmeier effect. 

So, we conclude that the next solar maximum will be of the similar amplitude as the previous one, or even something 
lower. 
This conclusion is based on the fact that the same method predicted a larger value for the maximum amplitude of the 24th solar cycle compared 
to the prediction for the 25th solar cycle, and taking into account the general lowering of the solar activity consistent with the Gleissberg cycle. 

This prediction is possible now when it is well established that the last solar minimum 
took place in December 2019, based on the smoothed monthly total (both solar hemispheres taken together) sunspot number. 

Finally, we successfully tested the modified minimum-maximum method to reproduce some of the earlier maxima in order to check the accuracy of the predictions. The maxima of solar cycles 21-24 were calculated using the previously available data. 
The 3 years before the minimum method is preferred over the simple minimum--maximum method, since it has better 
correlation coefficients and lower statistical errors. 

\section{Acknowledgments}
This work has been supported by the \fundingAgency{Croatian Science Foundation} under the project \fundingNumber{7549} "Millimeter and submillimeter observations of the solar chromosphere with ALMA". 
We acknowledge usage of the Sunspot data from the World Data Center SILSO, Royal Observatory of Belgium, Brussels. RB, AH, IS, and DS acknowledge the support from the \fundingAgency{Austrian-Croatian Bilateral Scientific Project} "Comparison of ALMA observations with MHD-simulations of coronal waves interacting with coronal holes". We would like to thank \v Zeljko Ivezi\'c for helpful comments and suggestions.

\appendix
\section{Additional material}

\begin{figure*}
\centering
\includegraphics[width=12cm]{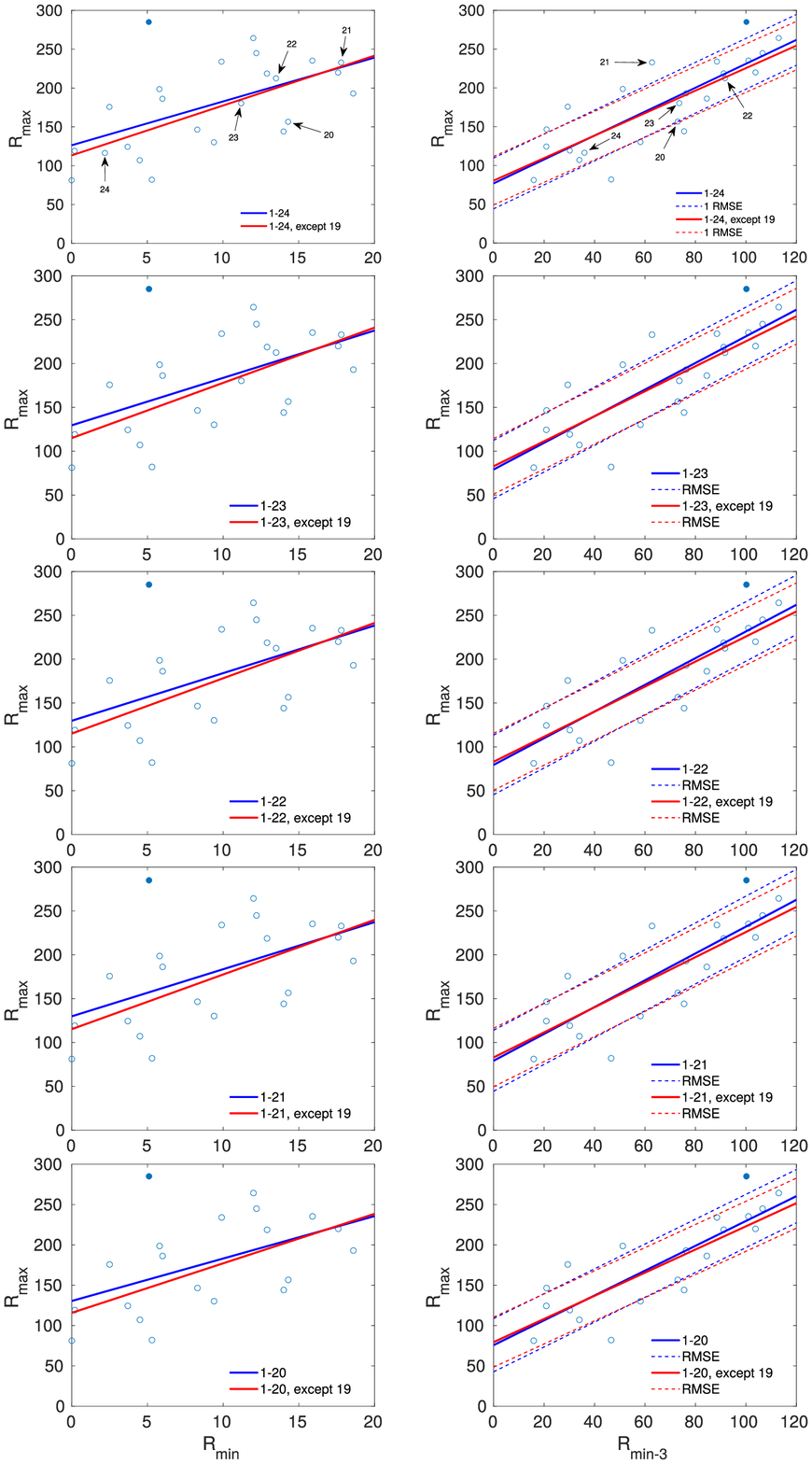}\par \caption{The peak smoothed monthly sunspot number in solar cycle maxima as a function
of the same quantity in the preceding solar minimum, for solar
cycles nos. 1-24 (first row), 1-23 (second row), 1-22 (third row), 1-21 (fourth row) and 1-20 (last row), for min--max method (left column) and min-3--max method (right column).
Least-square fits obtained with and without solar cycle no. 19 are presented with different colors, as indicated in the 
legend. The value for the solar cycle no. 19 is represented with the filled circle, while all other data points are 
represented with open circles.}
\label{Fig_4}
\end{figure*}

\bibliography{brajsa_sol_cyc_2021-v2}%

\end{document}